# Towards Intense Ultra-Broadband High Repetition Rate Terahertz Sources Based on Organic Crystals [Invited]

SAMIRA MANSOURZADEH*[1], TIM VOGEL[1], ALAN OMAR[1],
TOBIAS O. BUCHMANN[2], EDMUND J. R. KELLEHER[2], PETER U. JEPSEN[2]
AND CLARA J. SARACENO[1]

[1]*Photonics and Ultrafast Laser Science, Ruhr-University Bochum, Universitätsstraße 150, 44801 Bochum. Germany*
[2]*DTU Electro, Department of Electrical and Photonics Engineering, Technical University of Denmark, Ørsteds Plads 343, Kongens Lyngby, Denmark*

*\*Mansourzadeh.Samira@ruhr-uni-bochum.de*

**Abstract:** Increasing the average power of broadband, few-cycle terahertz (THz) sources is currently a topic of intense investigation, fueled by recent immense progress in high average power femtosecond laser driving sources at 1030 nm. However, many crucial applications would benefit not only from an increase in average power, but also from ultra-broad bandwidth, while maintaining high dynamic range at these frequencies. This calls for the challenging combination of high repetition rates and high average power simultaneously. Here, we discuss the recent progress in the promising approach enabled by organic crystals for THz-generation. Specifically, this review article discusses advances with the most commonly used organic crystals BNA, DAST, DSTMS, OH1 and HMQ-TMS. We place special emphasis on nonlinear and thermal properties and discuss future directions for this field.

## 1. Introduction

Few-cycle pulsed THz sources have become well-established tools in many fundamental fields of science and are being increasingly deployed in applied science and technology. In fundamental scientific research area, due to the overlap of THz radiation with a wide range of fundamental motions of ions, molecular networks, electrons, and electron spins in all phases of matter [1], THz radiation continues to be widely used in linear and nonlinear spectroscopy [2–4]. The potential of THz radiation has also been demonstrated in security [5], industrial non-destructive testing [6], and for medical diagnoses [7].

However, in these and many other fields, one challenge in studies with strong attenuation of THz waves, such as in aqueous/high humidity samples [8], is the typically low average power of pulsed THz sources e.g. less than milliwatt. Time-domain spectroscopy (TDS) partly circumvents the issue due to a high detection sensitivity; however, the typically required scanning over many probe pulses to reconstruct the time-traces and additional averaging over multiple pulses is challenging to combine with reasonably fast acquisition rates.

These challenges become significantly more severe for THz frequencies > 5 THz, yet an increasing number of fields are nowadays seeking ultra-broadband, sensitive TDS, for example for broadband characterization of emerging materials such as solar absorbers or metal oxide-based photoelectrodes [9,10]. In fact, the above-mentioned attenuation problem becomes more severe at higher frequencies, as materials of interest exhibit increasing THz absorption. Furthermore, detection sensitivity often rolls off at higher frequencies; as a result, typical commercially available TDS data extraction is restricted to < 5 THz due to low dynamic range (DR) at higher frequencies [11]. In electro-optic sampling (EOS) for example, a compromise needs to be met between bandwidth - which is in most cases constrained by velocity matching and calls for ultra-thin crystals - and electro-optic strength – which benefits from thick crystals.

Ultra-thin crystals are commonly used for high bandwidth detection but typically at the expense of DR. Whereas we exemplify the problem here with EOS, other detection methods, for example the THz air biased coherent detection (ABCD) [12], solid-state-biased coherent detection (SSBCD) [13], and photoconductive antennas [14], also have limiting constraints. This makes the need for high average power of even more critical importance for THz applications beyond 5 THz.

Traditionally, experiments calling for high average power were performed exclusively in accelerator facilities, which - although offering great parameter flexibility in terms of bandwidth and pulse properties [15]– are of restrictive access, strongly limiting experiments. In contrast, in the last few years, ultrafast laser driven sources with high THz average power can offer a table-top solution with wide accessibility. Presently, ultrafast laser systems based on the ytterbium (Yb)-ion have largely surpassed the kW-average power level [16–18], and even commercial systems are available with tens of watts to kilowatts of average power. Moreover, recent advancements in efficient pulse compression schemes [19–21] have overcome previous limitations regarding pulse duration, opening the door to broadband THz emission schemes. Based on this progress, THz sources using Yb-laser drivers have thus started to be demonstrated based on two-color filamentation in noble gases [22], and optical rectification (OR) [23,24]. The most explored route so far with high-average power 1030 nm driving lasers has been OR using nonlinear crystals exhibiting $\chi^{(2)}$ nonlinearity, mostly based on inorganic crystals. Among different available nonlinear crystals available for 1030 nm excitation, lithium niobate (LN) offers high nonlinear coefficients and high damage thresholds, and the tilted-pulse front (TPF) method allows for velocity matching and high conversion efficiency [25,26]. This has resulted in several breakthroughs with high-power excitation in various repetition rate ranges from tens of kHz to >10 MHz [27–29]. The highest average power so far demonstrated was recently reported in a conference contribution with 643 mW at 40 kHz using this technique and a high energy thin-disk amplifier system [30]. In spite of impressive results, the generated bandwidth is typically limited to below 2-3 THz in all cases, due to the high refractive index dispersion from near infrared (NIR) to THz range in LN and strong THz absorption due to phonon resonances which limit phase matching for higher THz frequencies. Another drawback of using LN is the tilted-pulse front geometry, which significantly complicates the experimental setup [31]. When considering semiconductor crystals, gallium phosphide (GaP) is also a promising candidate for high-power Yb-lasers due to favorable collinear velocity matching conditions at 1030 nm. This, in combination with phonon resonance shifted to higher frequencies, results in typically broader bandwidth compared to LN. Several results have been recently reported using GaP with Yb-lasers yielding multi-mW THz powers [32,33] and bandwidths up to 11 THz [33]; however, conversion efficiencies are limited to $\sim10^{-5}$ by multi-photon absorption due to the proximity of the excitation photon energy to the bandgap of the material at 2.26 eV [34].

One solution to strongly scale the THz average power and bandwidth simultaneously is the two-color gas plasma technique [35]. In this scheme, the achievable bandwidth is mostly limited by the driving laser pulse duration and not by phase matching constraints as in the nonlinear crystal-based schemes. Buldt *et al.* reported a high THz average power of 640 mW generated in a argon gas jet, at a repetition rate of 500 kHz [22], driven by a 16-channel Yb-fiber chirped pulse amplifier delivering 1.3 mJ of pulse energy, corresponding to 633 W of average power. The output of the driving laser is compressed to 37 fs using a multi-pass cell (MPC) compressor.

Moreover, it was found that the overall THz generation efficiency is of an order of magnitude larger in a three-color laser excited filament than the one produced by two-color pulses in an identical configuration [36]. Nevertheless, these plasma-based THz emitters suffer from a number of disadvantages, complicating their performance for applied fields. The main difficulty is the need for mJ-class femtosecond laser for efficient conversion, which again results in increased difficulty to implement this scheme at very high repetition rates.

An attractive alternative to high-power two-color plasma filaments and OR in inorganic materials are organic crystals, which combine the advantages of a broad bandwidth in a simple collinear velocity-matching geometry with high conversion efficiency [37]. During the past four decades, various types of organic nonlinear crystals have been developed exhibiting high macroscopic second-order optical nonlinearity. Impressive achievements have been realized with organic crystals: A high energy of 0.9 mJ [38], a high electric field of 6 MV/cm and an average power of 68 mW [39] and a record THz conversion efficiency of 6% [40]. However, these efforts focused mainly on achieving highest pulse energies at the expense of repetition rates, which makes the systems impractical for many applications. High-power Yb-lasers and their immense potential for operations at much higher repetition rates (> 100 kHz) could be a solution to overcome this problem, however, it remained unexplored with organic crystals until very recently. In fact, these crystals were believed to be unsuited for high average power and high repetition rate excitation due to their poor thermal properties. However, several recent results show that the thermal limitations can be circumvented to generate high-repetition rate, high average power, broadband THz-radiation, opening up promising new possibilities [41,42]. Figure 1 summarizes state-of-the-art of lab-based THz sources' performance, including OR in organic and inorganic materials and two-color plasma techniques. We note here we only highlight recent achievements based on 1030 nm excitation. It is evident that organic crystals occupy the space between watt-level average power and modest bandwidth from inorganic generators and the region of extreme bandwidth at generally modest average power generated by multi-color plasma sources.

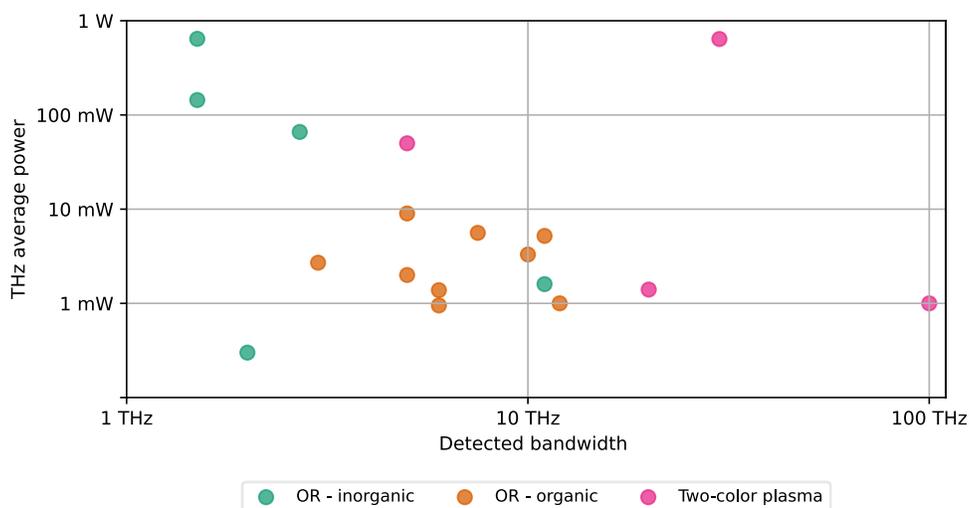

Fig. 1. State-of-the-art of lab-based THz sources: OR in organic and inorganic materials and two-color plasma [22,26,27,30,32,33,38,41–50].

The purpose of this review is to discuss these recent developments in broadband and efficient THz sources based on organic crystals excited with powerful 1030 nm femtosecond lasers and to provide the reader with a forward-looking perspective on future areas of research in this field. We start by summarizing the factors which affect the THz conversion efficiency in OR, followed by a detailed overview of the properties of some widely used organic crystals such as DAST (4-N,N-dimethylamino-4'-N'-methyl-stilbazolium tosylate), DSTMS (4-N,N-dimethylamino-4'-N'-methyl-stilbazolium 2,4,6-trimethylbenzenesulfonate), OH1 (2-[3-(4-hydroxystyryl)-5,5-dimethylcyclohex-2-enylidene] malononitrile), BNA (N-benzyl-2-methyl-4-nitroaniline) and HMQ-TMS (2-(4-hydroxy-3-methoxystyryl)-1-methilquinolinium 2,4,6-trimethylbenzenesulfonate), with a focus on missing parameters in the literature. We will discuss velocity matching conditions, compare their effective nonlinearity, coherence length and spectral bandwidth, as well as discuss the first conclusions on the influence of linear and nonlinear absorption, heating effects and damage on the THz conversion efficiency. In section 4, we summarize recent results in the area of high-power 1030 nm excitation and the current state-of-the-art. Finally, in section 5, we discuss future directions of research, including new types of organic crystals and perspectives to increase average power into the watt level regime.

## 2. Summary of important properties of optical rectification

Optical rectification is a well-known nonlinear effect occurring in non-centrosymmetric crystals with second order susceptibility, in which a low frequency component of the nonlinear polarizability acts as a low frequency (often called quasi-DC) source for THz radiation. We note that many textbooks and publications address the basic principles of OR for THz generation [51,52], therefore we focus here on providing a concise summary of the important properties and place our attention and discussion on relevant parameters in organic crystals and other factors that are crucial for average power scaling.

## 2.1 Nonlinear coefficient and velocity matching

In the femtosecond pump-pulse regime, OR can be viewed as a form of intra-pulse difference-frequency generation between frequency components of the driving pulse, generating a repetitive sequence of intrinsically phase-stable, transient electric field waveforms, whose bandwidth, under the assumption of perfect phase-matching and no spectrally dependent loss, is determined by the initial spectral extent of the pump pulse. As a result, to first approximation, the shorter the duration of the pump laser, the broader the bandwidth of the generated THz electric field. The build-up of this field over macroscopic propagation in the nonlinear medium is described by the inhomogeneous wave equation by considering induced nonlinear polarization by the pump pulse and the influence of the material dispersion. Therefore, the THz energy conversion efficiency ($\eta$) in the case of perfect velocity matching and neglecting pump absorption can be expressed by small-signal approximation based on the plane-wave model as follows [1,53]:

$$\eta = \frac{2d_{\text{eff}}^2 I_0 \omega_{\text{THz}}^2 L^2}{\varepsilon_0 c^3 n_{\text{gr,NIR}}^2 n_{\text{ph,THz}}} e^{\left(\frac{-\alpha_T L}{2}\right)} \frac{\sinh^2\left(\frac{\alpha_T L}{4}\right)}{\left(\frac{\alpha_T L}{4}\right)^2}, \tag{1}$$

where $d_{\text{eff}}$ is the effective nonlinear coefficient, $L$ is the crystal thickness, $\omega$ is the angular THz frequency, $\varepsilon_0$ is the vacuum permittivity, $I_0$ is the pump intensity, $n_{\text{gr,NIR}}$ is the group refractive index of the medium at the pump wavelength range, $n_{\text{ph,THz}}$ is the phase refractive index of the medium at THz frequencies, $\alpha_T$ is the THz absorption coefficient and c is the speed of the light in vacuum. Whereas considering no velocity mismatch and pump absorption is an oversimplification and does not apply in practice, this simple formula does provide insight into the parameters necessary to optimize the THz conversion efficiency. In, other factors are not considered in this equation which often limit efficiency, such as thermal effects, multi-photon absorption and free-carrier absorption. These effects can significantly affect the performance of the crystals for OR and are often not well-characterized in literature for a wide enough range of parameters.

Specifically, the effective nonlinear coefficient or $d_{\text{eff}}$ in pm/V is a crucial property of the nonlinear medium which gives the strength of the nonlinear polarization source. As shown in equation (1), the THz radiation efficiency is quadratically proportional to $d_{\text{eff}}$. The $d_{\text{eff}}$ of common materials is shown in Table 1 together with other parameters that will be discussed throughout this section. The notable advantage of organic crystals compared to inorganic crystals can be clearly seen in Table 1: large nonlinear coefficients and low dispersion of refractive index from NIR to THz regime which result in higher conversion efficiency and broader bandwidth. At the same time, it is evident that damage and thermal effects need to be tackled, which is the greatest challenge with high average driving powers and will be discussed in a following section.

Table 1. Overview of the relevant THz generation properties of different nonlinear materials. (*$n_{ph,THz}$ and $\alpha_T$ for DAST and DSTMS are given at 2 THz, due to a strong absorption peak at 1 THz)

| Material | $d_{eff}$ (pm/V) | $n_{gr,NIR}$ | $n_{ph,THz}$ @ 1 THz | $\alpha_T$ (cm$^{-1}$) @ 1 THz | Wavelength (nm) | Thermal conductivity (Wm$^{-1}$K$^{-1}$) |
|---|---|---|---|---|---|---|
| LN | 160 [37] | 2.22 [37] | 5 [37] | 17 [54] | 1000 | 5.7 [55] |
| GaP | 24 [37] | 3.36 [37] | 3.35 [37] | 0.2 [54] | 1000 | 77 [56] |
| DAST | 240 [37] | 2.8 [57] | 2.3* [58] | 10* [37] | 1030 | 1.58 [59] |
| DSTMS | 230 [37] | 2.4 [60] | 2.2* [60] | 28* [37] | 1030 | - |
| OH1 | 280 [37] | 2.5 [61] | 2.3 [61] | 0.1 [37] | 1030 | - |
| BNA | 234 [61] | 1.9 [62] | 2.05 [62] | 4 [63] | 1030 | 0.18 [64] |
| HMQ-TMS | 287 [65] | 2.17 [37] | 2.2 [37] | 47 [37] | 1030 | - |

Other than a large $d_{eff}$ a critical consideration for efficient THz generation is velocity matching. In fact, according to Eq. (1), increasing the nonlinear medium length significantly enhances efficiency, however, velocity matching needs to be fulfilled for power to build up: in the ideal case, the group velocity of the driving pulse should match the phase velocity of the generated THz wave in order to constructively interfere. In practice, dispersion of the driving and THz pulses do not allow for this condition to be fulfilled automatically. Different materials exhibit distinct ranges where the velocity matching condition is reasonably well satisfied. In Figure 2, we show the refractive index curves of commonly used organic crystals (DAST, DSTMS, OH1, BNA, and HMQ-TMS) in the NIR (green) and THz range (blue) as reported in literature. The vertical dashed line shows the wavelength of 1030 nm which is of interest for high power laser excitation. Due to the proximity of the phase index at THz frequencies and the group index in the NIR, organic crystals are an excellent candidate for efficient THz generation through OR. In the case of GaP, Figure 2(f), the proximity of the THz refractive index and group index of the pump pulse occurs in the vicinity of 1030 nm but for a significantly narrower frequency range than for organic crystals. We also show a similar curve for LN in Figure 2(g), illustrating the strong mismatch between $n_{gr,NIR}$ and $n_{ph,THz}$: in this case the tilted-pulse front scheme is required for power to buildup over mm-long lengths. It should be noted that there are discrepancies in the literature between the reported values on the refractive indices for some of the above-mentioned materials. Rather than inaccurate measurements, this may be due to differences in purity of the materials, generally making comparative conclusions difficult in the field.

When the velocity matching condition can be fulfilled collinearly, the experimental setup is greatly simplified, with the additional benefit of high spatial quality of the generated mode, allowing a small beam spot to be formed with a focusing optic to obtain a high electric field strength [1]. A practical metric to evaluate the nonlinear crystal length that is tolerable before velocity mismatch between THz and pump waves becomes too large is the coherence length ($L_c$), which is the length after which the mismatch is equal to $\pi$:

$$L_c = \frac{c}{2\nu_{THz}|n_{gr,NIR} - n_{ph,THz}|}, \tag{2}$$

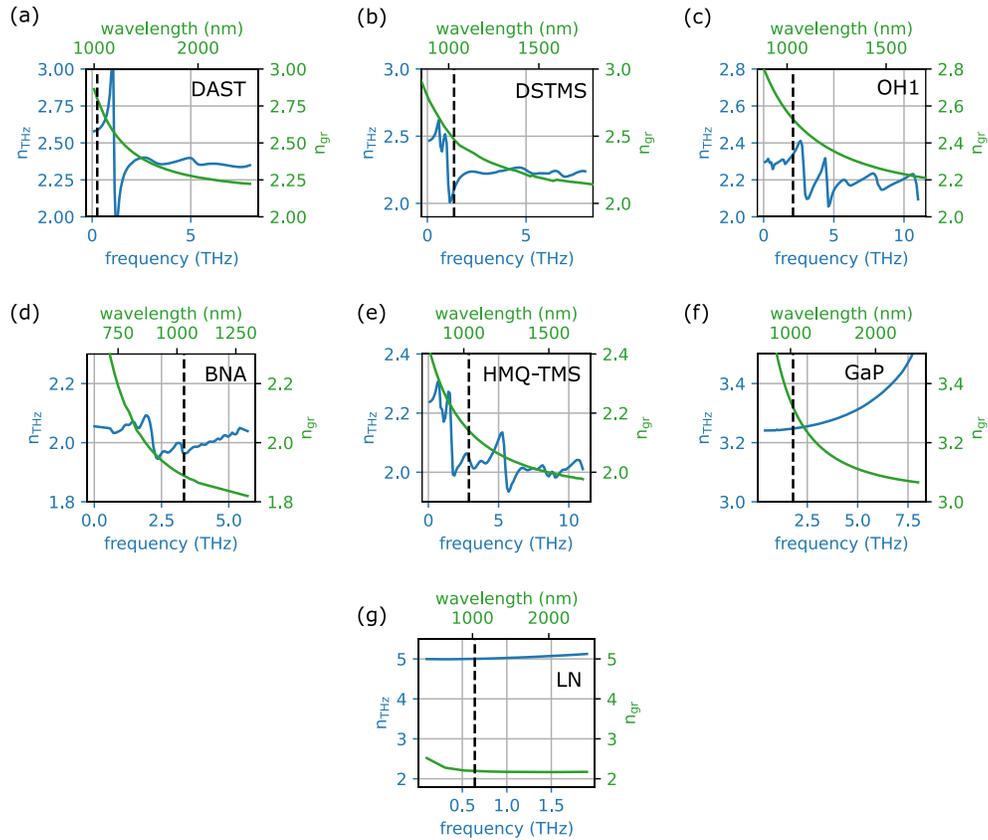

**Fig. 2.** THz refractive index and pump group refractive index for: a) DAST for the dielectric axis of $x_1$. $n_{\text{ph,THz}}$ redrawn with permission from [58] © The Optical Society. Data for $n_{\text{gr,NIR}}$ was taken from [57]. b) DSTMS for the dielectric axis of $x_1$ redrawn with permission from [60] © The Optical Society, c) OH1 for the dielectric axis of $x_3$ [61], d) BNA for the dielectric axis of $x_3$ redrawn with permission from [62] © The Optical Society, e) HMQ-TMS for the dielectric axis of $x_3$ [37] licensed by creative commons CC BY, f) GaP, $n_{\text{gr,NIR}}$ calculated from sellmeier equation from [66] and $n_{\text{ph,THz}}$ from [67] and g) LN in extraordinary axis, $n_{\text{ph,THz}}$ redrawn with permission from [68] © The Optical Society and $n_{\text{gr,NIR}}$ redraw with permission from [69] © The Optical Society. Please note that the wavelength range in each subfigure is different. The vertical dashed line shows the wavelength of 1030 nm which is of interest for high power laser excitation.

where, $v_{\text{THz}}$ is the THz frequency and $n_{\text{gr,NIR}}$ and $n_{\text{ph,THz}}$ have been defined before.

Figure 3 depicts the coherence length of the above-mentioned crystals (except LN) as a function of pump wavelength and THz frequency. In all cases, the 1030 nm wavelength is illustrated with a vertical dashed line, showing another reason why organic crystals were so far not explored in depth with 1030 nm wavelengths, as other pump wavelengths show better phase matching. In fact, most crystals show best velocity matching at wavelengths > 1100 nm, which are accessible with optical parametric amplifiers, but not with direct high-power Yb-lasers. For example, Figure 3(a) and Figure 3(b) show the coherence length of DAST and DSTMS with best velocity matching occurring in the vicinity of 1550 nm. It should be noted that at this pump wavelength, there is no high power (high energy) systems available. For the crystal OH1, velocity-matching cannot be fulfilled with a pump laser at a wavelength below ~1100 nm, and THz frequencies above 2 THz are only possible for pump wavelengths above 1500 nm. We note that another reason why organic crystals have not been explored in depth at 1030 nm is that Yb-lasers typically require external pulse compression to exploit the available bandwidth provided by organic crystals. This has been facilitated by efficient pulse compression schemes to bring these lasers to sub-100 fs with convenient, high efficiency setups only in recent years. Figure 3(d) shows the velocity matching for BNA with the most broadband operation at 850 nm. Significant bandwidth can also be obtained for 1030 nm excitation, but slightly thinner crystals are required to velocity match frequency components higher than 4 THz. In the case of HMQ-TMS, the optimal pump wavelength is above 1200 nm, however, also significant bandwidth can be obtained with slightly thinner crystals at 1030 nm. As can be seen in Figure 3(g), GaP is phase matched for a short range of pump wavelengths centered around 1030 nm, especially for the higher THz frequency components. Additionally, the low nonlinear coefficient of GaP makes the use of thin crystals result in a strong efficiency reduction.

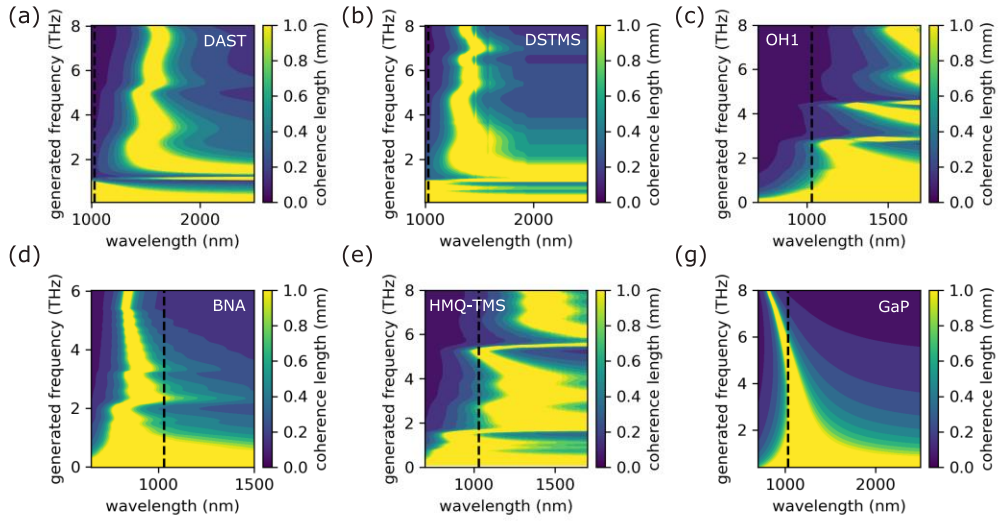

**Fig. 3.** Calculated coherence length versus generated THz frequency and pump wavelength for: a) DAST, b) DSTMS, c) OH1, d) BNA, e) HMQ-TMS and g) GaP. The refractive index of BNA at THz range was available only up to 6 THz.

In Figure (4) we plot the coherence length as a function of THz frequency at 1030 nm pump wavelength, which is the wavelength of interest for Yb-laser systems for the two most promising materials among these examples, BNA and HMQ-TMS. The figure shows the typical thickness of crystals required to velocity match up to a broadband frequency. Not surprisingly, these crystals have been the first ones to be explored at 1030 nm with powerful driving sources, as explained in more detail in the section below.

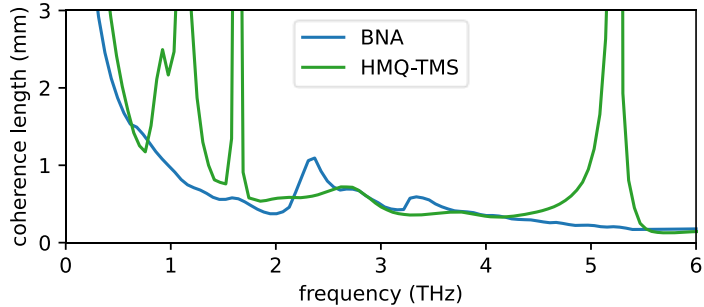

Fig.4. Coherence length cut at 1030 nm pump wavelength for: a) BNA (blue), and b) HMQ-TMS (green)

## 2.2 Linear absorption in organic crystals

Optimizing conversion efficiency in OR calls for minimal linear loss both at the driving wavelength and at the generated broadband THz frequencies. In the case of high driving average power, linear absorption can become an important source of temperature rise and can contribute to a reduction in efficiency. Table 2 shows the linear absorption coefficient ($\alpha_p$) of organic crystals in comparison with other common nonlinear materials such as GaP and LN at 1030 nm of pump wavelength.

Table 2. Linear absorption coefficient of selective organic and inorganic materials at the pump wavelength of 1030 nm and the averaged linear THz absorption in the range between 0 to 5 THz.

| Material | DAST | DSTMS | OH1 | BNA | HMQ-TMS | GaP | LN |
|---|---|---|---|---|---|---|---|
| $\alpha_p$ @ 1030 nm (cm$^{-1}$) | ~1 [70] | ~1 [71] | ~0 [72] | ~5 [62] | ~1 [73] | ~0 [74] | ~0.01 [75] |
| $\alpha_T$ averaged over 0-5 THz, (cm$^{-1}$) | 9.3 [37] | 6.4 [37] | 8.9 [37] | 5.6 [63] | 8.65 [37] | 4 [76] | 79 [77] |

As it can be seen here, LN, GaP and OH1 have a lower absorption coefficient at 1030 nm compared to other materials. Among organic crystals, BNA has the highest absorption coefficient of 5 cm$^{-1}$.

Another crucial parameter which limits the THz conversion efficiency and bandwidth is THz absorption in crystals. It is mainly caused by lattice vibrations or optical phonon resonances, leading to excess absorption. Each nonlinear material used for OR has phonon resonances at specific frequencies, causing loss in the vicinity of these frequencies. In the organic crystals, the strong optical phonon resonance in DAST occurs near 1.1 THz. In OH1, the strongest resonance occurs at a frequency above 10 THz, with two other peaks with smaller amplitude at about 3 THz and 4.5 THz [37]. In order to estimate the THz absorption in organic crystals in comparison to inorganic materials, the average THz absorption coefficient over 0-5 THz ($\alpha_T$) for different materials can be calculated by integrating the absorption coefficient curves over the THz frequencies. The curve for DAST, DSTMS, OH1 and HMQ-TMS are taken from [37] and for BNA from [63]. This value for DAST is 9.3 cm$^{-1}$, DSTMS is 6.4 cm$^{-1}$, OH1 is 8.9 cm$^{-1}$, BNA is 5.6 cm$^{-1}$ and HMQ-TMS is 8.65 cm$^{-1}$ (summarized in the Table 2). As a reference, GaP has a similar value to organic crystals and it is about 4 cm$^{-1}$ [76]. These values for organic crystals are about one order of magnitude lower than the one for LN calculated to be approximately 79 cm$^{-1}$ [77] up to 4 THz. The lower THz absorption in organic crystals is one of the main factors contributing to their broadband and high efficiency capabilities.

## 2.3 Additional factors limiting the efficiency: nonlinear absorption and thermal effects

In many inorganic crystals it is well known that multi-photon absorption (MPA) is the main limitation for efficiency scaling via increasing the intensity on the crystal [78]. MPA produces free carriers, which can absorb the generated THz radiation through free carrier absorption (FCA). This process reduces the efficiency of THz generation via OR and additionally reduces the damage threshold of the nonlinear material [79]. Simultaneous absorption of multiple photons occurs if the sum of the energy of these photons is at least equal to the bandgap energy of the nonlinear material. Therefore, materials with a higher bandgap energy are less likely to absorb multiple photons than media with a lower bandgap energy. In order to quantify MPA, the bandgap energy of selective organic and inorganic materials is given in Table 3. Concerning BNA, using the absorption cut-off wavelength which is about 500 nm [61], the bandgap energy can be estimated about 2.5 eV which is comparable to the bandgap of other organic crystals.

Among inorganic materials, LN does not suffer from two and three photon absorption due to the high bandgap energy of 3.7 eV.

Table 3. Bandgap energy of selective organic and inorganic materials.

| Material | DAST | DSTMS | OH1 | BNA | HMQ-TMS | GaP | LN |
| --- | --- | --- | --- | --- | --- | --- | --- |
| Bandgap (eV) | 2.19 [80] | 2.28 [81] | 2.17 [82] | - | 2.95 [82] | 2.26 [83] | 3.7 [84] |

In the context of high average power lasers, another critical consideration are thermal effects and corresponding changes in material properties. When average power levels on the crystal exceed several watts or even more, linear absorption can become a significant part of total absorption. This is particularly severe at high repetition rates, where focusing needs to be appropriately downsized to reach high intensity at moderate pulse energy. Additionally, MPA also results in heating. The relative contributions to the temperature-rise in the crystal is strongly dependent on the repetition rate and pulse peak intensity available for the experiments. Disentangling their relative contributions and understanding in detail the effects of temperature on material properties is critical in understanding how to scale OR to high average powers and these aspects have so far been nearly unexplored in the context of OR. Many incomplete and even contradictory findings exist in the literature. In LN thermal effects are mentioned in [27] as a cause of efficiency reduction, however, no detailed study or experimental evidence is provided. In [31], it is suggested that in the high repetition rate regime, even with tight focusing, walk-off rather than thermal effects are the limiting factor, somewhat in contradiction with [27] which used lower average pump intensities. In [85], the origin of thermal load in GaP was explored and it was concluded that even at 100 W average power and 13 MHz repetition rate, MPA was the main limiting factor and linear absorption had little influence on heating. Additionally, the change of THz refractive index was measured as a function of temperature showing no significant change in velocity matching conditions in a wide temperature (T) range (from 77 K to 500 K).

In organic crystals, thermal properties are significantly worse than for inorganic crystals used for OR (see Table 1), therefore such explorations are even more critical and strongly dependent on the targeted repetition rate. In [49], a detailed study of the thermal properties of BNA under high-power pumping and MHz repetition rate was reported and it was shown that thermal effects and thermal damage represent the main limitation to conversion efficiency. Heat removal using an optical chopper at lower duty cycles and a diamond heatsink were critical to reach reasonable conversion efficiencies.

Besides passive cooling with a diamond or sapphire substrate, active cooling of the crystal using a cryostat helps to improve the performance of THz generation in organic crystals, however, more detailed exploration is needed for a full understanding of thermal properties; in fact, the crystals having poor thermal conductivity, active cooling needs careful implementation to be effective. In [86], the temperature-dependent optical properties of BNA at THz frequencies in a temperature range from 80 K up to 300 K were investigated. The results showed a negligible change in refractive index, but a significant reduction in THz absorption of −24% was achieved by cooling the crystal from room temperature down to 80 K. Moreover, this study showed cryogenically cooled OR experiments in BNA pumped with a thin-disk laser, operating at ~13 MHz repetition rate in a first attempt to correlate the observed trends with the measured crystal properties. An increased THz field strength (+23%) was observed, most likely corresponding to a reduced THz absorption at low temperatures. In that experiment, no significant improvement in the damage threshold of BNA was observed, possibly due to a poor heat dissipation of the bare crystal (unbonded on a sapphire substrate) used in that experiment. These few experiments show that more experimental work on elucidating the damage mechanisms at high average power and high repetition rate are required.

## 2.4 Damage thresholds

Damage thresholds are critical to understand scaling of any nonlinear process and represent the ultimate limit when increasing the pump intensity on the crystal. The causes of damage threshold are complex to disentangle and vary with different pulse parameters. We aim here to consolidate existing literature on damage thresholds measured with fs-pulsed systems, which is of critical importance to average power scaling.

In [87], the damage threshold of DAST was investigated as a function of pulse repetition rate up to 100 Hz. The authors demonstrated that the thermal diffusion in the crystal at repetition rates of more than 50 Hz is limited by the relaxation time of DAST. In [88], the damage threshold of DAST and DSTMS was investigated with a laser pulse at a central wavelength of 800 nm, repetition rate of 100 Hz and pulse duration of 60 fs. A damage threshold of 300 $GW/cm^2$ (20 $mJ/cm^2$) was measured for DAST and a slightly lower value was found for DSTMS. By using a pump wavelength of 3.9 µm, the optical damage threshold of DAST was increased to 110 $mJ/cm^2$, which can be attributed to a suppression of two photon absorption (TPA) [40].

The damage threshold of HMQ-TMS was reported to be approximately 1.8 $mJ/cm^2$ when pumped with a 1030 nm laser at repetition rate of 10 MHz [89], which is much lower the damage threshold at 100 Hz (>20 $mJ/cm^2$) reported in [90]. The high repetition rate causes the dominant damage mechanism to become mostly thermal at a lower fluence, rather than governed by the high peak intensity of a low repetition rate pump pulse [89].

For the organic crystal BNA, the damage threshold reported in [91] is more than 6 $mJ/cm^2$. The crystal in this experiment was pumped with an 800-nm wavelength laser at a repetition rate of 100 Hz and pulse duration of 50 fs. Another study on damage threshold of BNA is given in [64], where the crystal was pumped with a laser with central wavelength of 800 nm and repetition rate of 500 Hz. The reported damage threshold in this case was 4.8 $mJ/cm^2$ in the crystal with thickness of more than 300 µm. At the higher repetition rate of 540 kHz and a pump wavelength of 1030 nm in [42], the damage threshold of BNA was reported to be approximately 1.7 $mJ/cm^2$. The lower damage threshold at higher repetition rate can be rationalized with the detailed investigation of thermal effects reported in [49] where it was shown that a trade-off between the repetition rate and conversion efficiency needs to be met due to thermal effects and the correspondingly reduced damage threshold. All damage threshold data from previously mentioned experiments are summarized in table 4.

Table 4. Damage threshold of organic crystals at different pump repetition rate and wavelength.

| Material | Damage threshold ($mJ/cm^2$) | Damage threshold ($GW/cm^2$) | Repetition rate | Pump wavelength (nm) |
|---|---|---|---|---|
| DAST [88] | 20 | 300 | 50 Hz | 800 |
| DAST [40] | 110 | 968 | 20 Hz | 3900 |
| BNA [91] | 6 | 106 | 100 Hz | 800 |
| BNA [42] | 1.7 | 33 | 540 kHz | 1030 |
| BNA [64] | 4.8 | 42 | 500 Hz | 800 |
| HMQ-TMS [90] | 20 | 352 | 100 Hz | 1500 |
| HMQ-TMS [89] | 1.8 | 53 | 10 MHz | 1030 |

We note that these values are to be treated with caution when comparing damage thresholds at very high repetition rates. In fact, when the time between pulses becomes significantly shorter than the thermal relaxation time of the crystal, the low repetition rate 'burst' energy (caused by the sum of the fs-pulses during the chopper wheel cycle open time rather than the single pulse energy) is the correct metric to calculate the damage threshold. In [49], it was shown that by considering the burst energy, rather than the single pulse energy, the damage thresholds can be compared closely to previous literature using low repetition rate systems.

## 3. State-of-the-art and recent developments using 1030 nm high-power excitation

For broadband OR in organic crystals, highest efficiencies are achieved with maximum peak intensity and optimized central wavelength, as discussed in the previous section. In addition to boosting the efficiency, short pulses and a correspondingly broad spectrum are required to support the generation of the broadest THz bandwidths. For applications of these systems, a high pulse repetition rate is desired to accelerate measurement times in potentially complex target experiments. These considerations lead to a general prescription for the desired parameters of the driving laser system: a central wavelength to achieve broad velocity matching, high peak power, short pulse duration and high average power.

Historically, organic crystals were first tested with high energy optical parametric amplifier systems with tunable central wavelength typically between 1250 nm and 1500 nm [62,92], and specialized Cr:forsterite systems to access best excitation wavelengths >>1000 nm as discussed in the previous section. This resulted in a multiplicity of record holding systems, mostly operating at low repetition rate < 1 kHz. For example, DAST, DSTMS and OH1 at 1250 nm using a high energy optical parametric amplifier were tested which resulted in a conversion efficiency of more than 3.3% in OH1, >2.2% in DAST and about 1% in DSTMS [47]. More recently, organic crystals such as BNA were presented with suitable velocity matching for lasers with a central wavelength of 800 nm, which also resulted in many advances pumped directly with Ti:Sa lasers, for example efficiency of 0.5% at repetition rate of 1 kHz [64] and efficiency of 0.25% at repetition rate of 100 Hz [91].

However, all the above-mentioned driving laser systems are limited in average power and most importantly, repetition rate, to typically <5 W and <1 kHz. Modern Yb-lasers with a central wavelength of 1030 nm capable of reaching equivalent pulse energies with orders of magnitude higher repetition rate started to be used to pump organic crystals in 2020, with a first demonstration of sub-mW of THz power at repetition rate of 10 MHz using HMQ-TMS [89] followed by [41]. Afterwards, BNA was investigated at 13.3 MHz in [49] and at 540 kHz of repetition rate in [42].

Despite this potential to increase the available repetition rates, Yb-doped laser systems have remained relatively unexplored for THz generation via OR in organic crystals. On the one hand this is due to the typically non-optimal velocity matching conditions at 1030 nm for most crystals as discussed in the previous section. Furthermore, their thermal properties were believed to be a limiting factor for high average power excitation. Last but not least, Yb-doped high-power lasers suffered for very long from long pulse durations and difficulties to reach sufficiently short pulse durations. As discussed above, this limits the optical to THz conversion efficiency and bandwidth.

For many years, hollow-core capillaries were the mainstream solution for pulse compression of high peak and average power laser systems [93]; however, they are typically lossy and impractical to use due to extreme bending loss. Hollow-core photonic crystal fibers were adopted for higher repetition rates as well but were generally impractical for hundreds of watts of average power [94,95]. Progress in the development of new techniques in compressing the pulse externally using Herriott-type MPC based compressors [21] represented a breakthrough in this area. MPCs have proven to be an efficient technique for free-space spectral broadening and compression with high throughput of more than 90% and great flexibility in terms of input parameters and wavelengths [20,96–98]. The basic principle behind MPCs is to prevent beam degradation by dividing nonlinear spectral broadening into small steps with sufficient free-space propagation between them, thereby suppressing spatiotemporal couplings [99]. Moreover, these systems achieve high compression ratios without compromising the overall optical transmission and beam quality. For instance, compression factor of over 32 is achieved in gas-filled MPC with an $M^2$ of 1.16×1.19 [100]. In this way, it is possible to combine the advantages of free-space propagation - which is crucial for lasers with high average powers - while still preserving excellent beam quality, even at large broadening factors after several passes through a solid material [21].

Besides near infrared lasers, organic crystals have also started to be explored with short-wave mid-infrared laser system at 1950 nm and 3000 nm [40] reaching remarkably high efficiency of ~6% and setting a new direction for future developments. At much higher repetition rates, the use of erbium-doped fiber lasers with central wavelength of 1560 nm [101] has also recently been reported demonstrating a 15-THz bandwidth source based PMB-4TFS (2-(4-(4-(hydroxymethyl) piperidin-1-yl)styryl)-3-methylbenzothiazol-3-ium 4-(trifluorome-thyl) benzenesulfonate), however at moderate average power levels.

### 4. State-of-the-art in 1030 nm excited high-power organic crystal THz sources

In this section, an overview of studies on high power (mW-class) and broadband THz sources (> 5 THz) based on nonlinear organic crystals pumped with high power, high repetition rate Yb-based lasers is presented.

### 4.1 Milliwatt-class THz sources based on HMQ-TMS

In [41], Buchmann at al. demonstrated milliwatt-level, few-cycle THz pulse generation at 10 MHz repetition rate. The Yb-doped fiber laser with an external pulse compression based on a large mode area solid-core photonic crystal fiber is used to pump HMQ-TMS. The pump pulse duration after compression stage was 22 fs. THz generation was explored in two distinct thicknesses of the organic crystal, 0.25 mm and 0.45 mm leading to a maximum conversion efficiency of ~0.055%, an order of magnitude higher than that achieved with inorganic nonlinear crystals such as GaP, for similar pump parameters i.e., wavelength and repetition rate. A THz average power of 1.38 mW at pump power of 2.5 W was measured using a calibrated pyroelectric power meter. The achieved THz electric field was characterized using an EOS setup. A 0.3 mm GaP crystal was used to detect the THz trace. The spectrum after propagation through ~45 cm of air at ~25% relative humidity has a broad bandwidth extending up to 6 THz as it is shown in Figure 5. The focused THz beam had a peak field strength ~6.4 kV/cm. This moderately strong-field THz source would be well suited to a variety of applications in ultrafast THz spectroscopy. Previously, in their work in [89], as was observed with all organic crystals, HMQ-TMS possesses a low damage threshold (1.8–3.6 mJ/cm² for MHz operation).

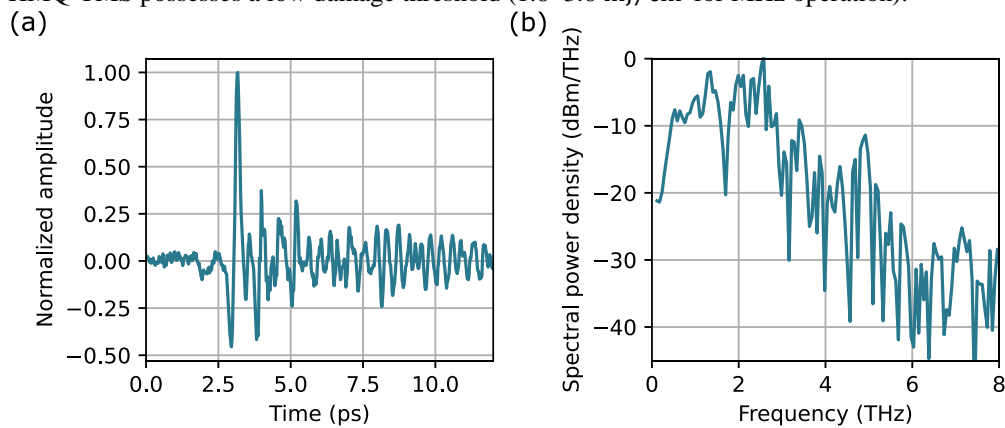

Fig. 5. a) Temporal waveform of the generated THz electric field emitted from HMQ-TMS. b) Output spectral power density. Licensed by creative commons CC BY [41].

This crystal can be prepared as large-area crystals by a simple cleaving method [102], to increase the pump beam area at high average power and stay below the damage threshold fluence. It should be noted that the conversion efficiency here is 5.5 times higher than the results reported in [89] where the peak power is lower due to longer pump pulse duration (30 fs instead of 22 fs). Also, additional increase in THz power was limited due to the solid-core nature of the compression setup which suffered from self-focusing at ~2MW of peak power. Hence, further THz enhancement can be expected when pumping HMQ-TMS using Herriott-cell based compression setups.

### 4.2 Milliwatt-class THz sources based on BNA at MHz repetition rate

In [49], OR in a diamond-heatsinked organic crystal BNA pumped by an externally compressed high-average power mode-locked thin-disk oscillator at MHz repetition rate was reported and a detailed investigation of thermal behavior of BNA crystals was demonstrated. By disentangling the effect of pump power and pump energy using an optical chopper with adjustable duty cycle, it was shown that the saturation effect in the conversion efficiency is only due to the pump power and not the pulse energy, which confirmed that the main limiting effect in this high average power, high repetition rate regime are thermal effects.

Additionally, it was confirmed that the highest conversion efficiency is achieved with the smallest duty cycle, since for this configuration, the pulse energy and the peak intensity are highest and thermal effects that decrease the conversion efficiency or lead to damage are reduced. In the optimized condition with the lowest possible duty cycle of the chopper (10%), a maximum THz power of 0.95 mW with a smooth spectrum extending beyond 6 THz with a spectral dynamic range of more than 50 dB was achieved, see Figure 6. It should be noted that a 0.2 mm GaP is used as a detection crystal in the EOS setup.

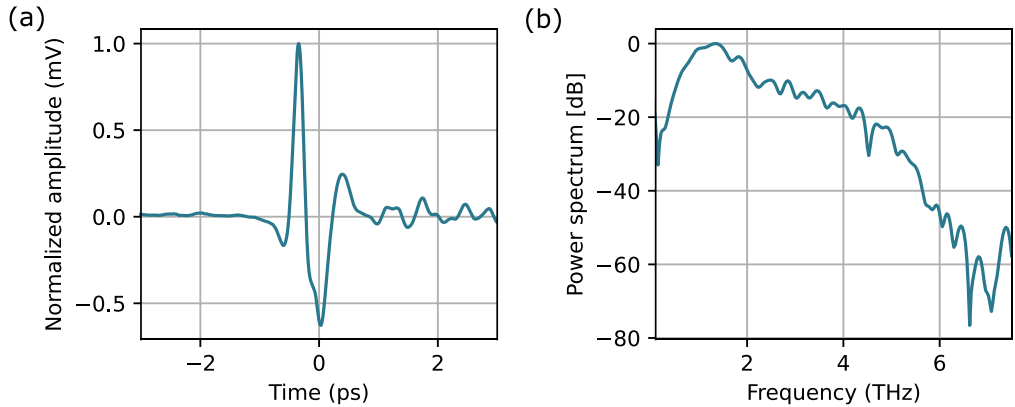

Fig 6. a) Temporal waveform of the generated THz electric field emitted from BNA. b) Corresponding spectrum calculated using Fourier transform [49].

### 4.3 Towards strong-field broadband THz sources at multi-100 kHz rep rates

BNA was also investigated in [42] using a commercial Yb-based laser. The laser was operated at a repetition rate of 540 kHz and a central wavelength of 1030 nm. The initial pulse duration of 240 fs from the laser was temporally compressed using a Herriott-type MPC down to 45 fs. In order to control the thermal load on the crystal at the high repetition rate of 540 kHz and avoid the damage of the crystal, an optical chopper with a duty cycle of 50% was placed before the crystal. The THz average power was directly measured using a calibrated THz power meter. In the optimized condition, the crystal was pumped up to 4.7 W without any irreversible damage and a maximum THz average power of 5.6 mW with an efficiency of 0.12% was reported. The THz electric field was characterized using a 0.2 mm GaP crystal in a standard EOS setup. The results are shown in Figure 7 in both time and frequency domain. The spectrum has a wide bandwidth that spans up to 7.5 THz. This source had a peak electric field of 29 kV/cm.

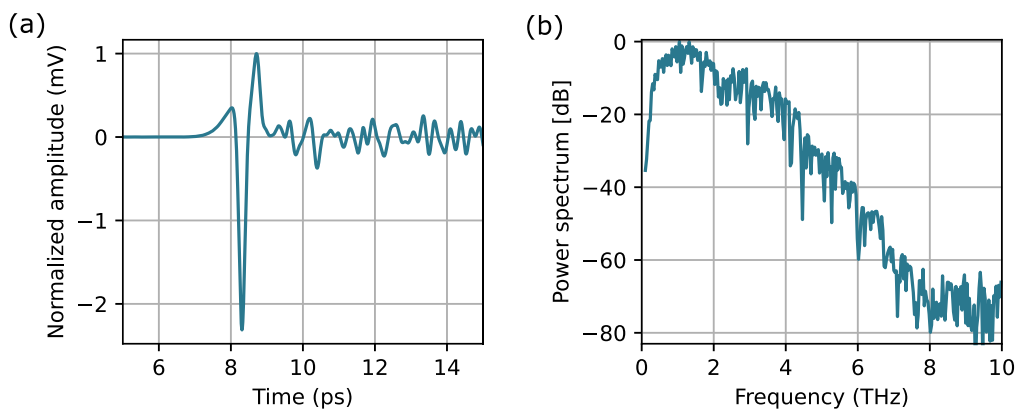

Fig. 7 a) Temporal waveform of the generated THz electric field emitted from BNA. b) Corresponding spectrum calculated using Fourier transform. Licensed by creative commons CC BY [42].

## 5. Future directions

In this section, we discuss several aspects about the future direction of organic crystals excited with high average power Yb-lasers. These aspects will help to improve the performance of THz sources based on organic crystals in terms of power, bandwidth and reliability.

### 5.1 Further understanding of thermal effects and advanced cooling geometry

From the above described results, it becomes apparent that the conversion efficiencies reached at high repetition rates are lower than record values obtained at <1 kHz. Whereas recent studies about efficiency limitation in scaling of organic crystals to high repetition rate and high average power pumping in [49] clearly indicated that heat effects were correlated with the reduction in THz conversion efficiency, it was not possible to precisely identify the physical mechanism of the generated heat. Further investigations on thermal behavior of the crystal is required. In fact, as discussed previously in the case of high average powers and high repetition rates, a significant amount of heat can be generated due to pump laser absorption. Both linear and nonlinear absorption of the pump laser can result in a large temperature rise in the crystal, shown by strong heat accumulation. Therefore, it is required to quantify heat power generation due to these absorption mechanisms in organic crystals, for example in the same way as in [85] using a cryostat and a temperature controller. In order to distinguish between linear and nonlinear absorption, the experiment was conducted with continuous wave and ultrafast pulsed lasers individually, and in each case the generated heat was measured. For organic crystals, the same procedure will be critical to elucidate limiting mechanisms.

A critical factor to improve the THz generation performance will be using efficient cooling techniques. As mentioned before, dissipating heat using a high-conductive substrate such as sapphire [103] or diamond [49] increases the damage threshold of organic crystals. A sapphire substrate was found to increase the damage threshold, thermal stability and also the conversion efficiency by a factor of 4 [103]. One further improvement is to sandwich the crystal between two heatsinks. It should be noted that one of the heatsinks should be transparent in the THz range to be able to couple out the radiation from the crystal. Other solutions such as using the crystal in a back-reflection geometry [104], could be another route to explore. Combining these cooling geometries with active cooling such as demonstrated in [88], could result in significant improvements. However, according to observations in [105], changing the temperature of the crystal actively (300 K down to 80 K) changes the structure of BNA (and presumably other organic crystals) resulting in changes in non-negligible changes in absorption and refractive index, which need to be carefully factored in. Additional studies of temperature-dependent properties of organic crystals, such as the one performed in [88], in particular at broader bandwidths, will be an important ingredient for future steps forward.

### 5.2 Improving crystal quality and repeatability of samples

Although intense and broadband THz sources based on organic crystals are attractive for users in many applications such as spectroscopy, day-to-day reproducibility of the experimental conditions and stability of the THz radiation in these sources remains challenging due to e.g., crystal quality degradation over time and variability of the crystals. Moreover, the crystal quality and size have a large effect on the THz generation process to achieve high power radiation. In this regards, a study reported in [106] demonstrated that low crystal quality reduces the THz output and the production of larger, higher quality crystals promises a significant gain in THz generation combined with a broad generation spectrum. High-quality organic crystal plates can be grown using recently developed techniques reported in [62,107,108]. The quality of the crystals becomes critically important when they are used in the THz detection scheme as well, which will be briefly discussed in the next section.

### 5.3 Broadband organic crystal as detectors

High quality organic crystals can be used to detect THz radiation in a broad spectral range. Using the identical crystal type for generation and detection will automatically give the largest possible detectable bandwidth. However, typical organic crystals such as DAST, DSTMS and OH1 can hardly be used in classic EOS detection scheme due to their high natural birefringence. In these cases, the orthogonal field components lose their coherence after propagation through the detection crystal which cannot be compensated [9]. There are multiple schemes to overcome this problem. One of the most propitious techniques is the so-called THz induced lensing effect based on focusing and defocusing of the ultrafast probe pulse. In this technique changes to the probe beam profile in the presence of the THz electric field is detected, allowing the use of the highly birefringence materials with high electro-optic coefficients for detection [109]. Another technique allowing to use above-mentioned organic crystals as a detection material is based on laser pulse energy changes introduced in [110]. It is based on the changes in the total laser pulse energy of the orthogonal polarization components of the probe beam induced by THz waves in the detection crystal. Further broadband, high DR detection techniques are a natural complementary research area of the above discussed high-power THz sources.

### 5.4 New organic crystals

Research and development of advanced organic materials that possess useful properties is vital for the advancement of a number of fields, such as nonlinear optics, catalysis, and electron transport. However, development of new organic materials for intense THz generation is a complex process involving a combination of molecular design, crystal structure determination, and growth optimization [108,111,112]. Valdivia-Berroeta *et al.* [108] accelerated this development process by employing structural data mining techniques to identify organic materials with ideal non-centrosymmetric crystal properties from the Cambridge Structural Database. Following this, the authors performed a first-principles investigation of the molecular property that is most crucial to nonlinear optical properties i.e. hyperpolarizability, and they inspected crystal packing to rank the molecular crystals based on their potential for generating THz. Through this data mining, they discovered, synthesized and characterized a number of new organic THz generation crystals that exceed the performance of industry standards in terms of crystal size and quality. Some of these new organic materials are PNPA ((E)-4-((4-nitrobenzylidene) amino)-N-phenylaniline), NMBA ((E)-4-((4-methylbenzylidene)amino)-N-phenylaniline) and MNA (2-Amino-5-Nitrotoluene). One example of recent developments in this area is with the promising organic crystal MNA. MNA originally was identified about 40 years ago as a powerful material for nonlinear applications due to its high molecular hyperpolarizability, favorable non-centrosymmetric crystal packing, and relatively large molecular number density. The nonlinear coefficient of MNA was reported to be 250 pm/V [113] which is comparable to the other organic crystals given in Table 1. However, it has not been widely used to generate THz radiation due to the difficulties in synthesizing sufficiently large crystal sizes. Very recently, Palmer *et al*. [107] pioneered a breakthrough method to consistently grow single, large, and high quality MNA crystal which was unachievable over the past 40 years by a two-step crystallization process. The maximum THz generation efficiency by measuring the THz pulse energy with a pyroelectric detector at 1250 nm wavelength was 3% at a maximum fluence of 1.3 mJ/cm$^2$. It was demonstrated that the THz generation efficiency of MNA crystals is comparable or exceeds the efficiency of the commonly used crystal BNA, which is a derivative of MNA. These studies confirmed that MNA is an excellent and readily available candidate for high intensity and broad THz generation, as MNA is commercially available and does not require the additional synthetic steps for synthesizing BNA. Rader *et al*. [114] characterized the THz generation in PNPA and compared the results with the commercially available DAST and OH1 crystals. A THz peak electric field of 2.9 kV/cm was achieved using PNPA, which was higher than in DAST (2.2 kV/cm) and OH1 (1.5 kV/cm) measured at their maximum applicable fluence conditions. Furthermore, PNPA generated a broad and smooth spectrum extending to 5 THz. In general, new organic crystals might open new options for efficient, broadband, and high-power 1030 nm pumping. Table 5 shows some of the relevant parameters of MNA and PNPA crystals, which were taken from first THz generation experiments:

Table 5. Nonlinear coefficient, refractive index, and absorption coefficient in the THz regime for MNA and PNPA.

| Material | $d_{eff}$ | $n_{ph,THz}$ @ 1 THz | $\alpha_T$ (cm$^{-1}$) @ 1 THz |
|---|---|---|---|
| MNA | 250 [1] | 2.5 [2] | ~2 [2] |
| PNPA | - | 2.44 [3] | ~10 [3] |

## 6. Conclusion and outlook

In recent years, organic crystals have attracted a great deal of attention within the rapidly growing field of high-power THz source development. These crystals provide a promising platform to achieve broadband high-power THz sources at high repetition rate in a simple collinear geometry. In this paper, we reviewed the unique properties of these crystals that provide an opportunity to not only achieve high THz power at high repetition rates but also an ultra-broad THz spectrum. We consolidated properties of these crystals from literature which affect the THz generation process based on OR such as the nonlinear electro-optic coefficient, the coherence length, THz absorption and multi-photon absorption, all with a particular emphasis on parameters at 1030 nm. Operation in "burst mode" utilizing an optical chopper wheel with the pump laser is the key point to reach the high efficiency in the high power and high repetition rate of excitation. We note that burst-mode operation requires even higher average power front-ends, thus recent laser developments will be a strong supporter of this area.

Current results based on BNA and HMQ-TMS open a variety of potential future directions. More detailed temperature dependent studies of relevant crystal parameters are required for complete understanding of the limiting mechanisms at high average power pumping and high average intensities on the crystals, in particular with the goal of harnessing temperature-rise and damage. More efficient cooling geometries and active cooling schemes will be key to apply higher average powers and intensities to reach higher efficiencies at high repetition rate. The pump laser systems themselves can be further optimized to achieve higher THz power, efficiency, and bandwidth; for example, a 100W-class MHz repetition rate sub-10 fs laser system would be highly attractive and is clearly within reasonable reach [115] and will result in higher power and also broader THz bandwidths. Last but not least, engineering of these crystals and rapid identification of new crystals with higher nonlinearity using new approaches and data mining techniques will support further improvements. We are confident that continued THz power-scaling deep into the hundreds of mW range and towards the Watt-level at MHz repetitions rates is possible by careful optimization of the aspects discussed in this review.

**Acknowledgements.** This project was funded by the Deutsche Forschungsgemeinschaft (DFG) under Germany's Excellence Strategy - EXC 2033 - 390677874 – RESOLV and also under Project-ID 287022738 TRR 196 (SFB/TRR MARIE). These results are part of a project that has received funding from the European Research Council (ERC) under the European Union's Horizon 2020 research and innovation programme (grant agreement No. 805202 - Project Teraqua). The project "terahertz.NRW" is receiving funding from the programme "Netzwerke 2021", an initiative of the Ministry of Culture and Science of the State of Northrhine Westphalia. Additionally, we acknowledge the support by the MERCUR Kooperation project "Towards an UA Ruhr ultrafast laser science center: tailored fs-XUV beam line for photoemission spectroscopy.

**Disclosures.** The authors declare no conflicts of interest.

**Data availability.** Data underlying the results presented in this paper are not publicly available at this time but may be obtained from the authors upon reasonable request.